# Impact of two-level fuzzy cluster head selection model for wireless sensor network: An Energy efficient approach in remote monitoring scenarios

P.K. Dutta[1], M.K.Naskar[1], O.P.Mishra[2]

[1] *Electronics and Tele-Communication Dept. Jadavpur University, Kolkata, West Bengal , India*
*ascendent1@gmail.com*

[1]*Electronics and Tele-Communication Dept. Jadavpur University, Kolkata, West Bengal , India*
*mrinalnaskar@yahoo.co.in*

[2] *SAARC Disaster Management Centre (SDMC), IIPA Campus, Ring Road, Delhi*
*opmishra2010.saarc@gmail.com*

Abstract

*The robust application of wireless sensor networks has increased during the past decade due to the potential use of wireless nodes in transmission of information by decreasing latency for surveillance and monitoring. The study proposes an Energy Efficient Dynamic Scenario (EEDS) for cluster head allocation for optimum balance in the energy consumption of the whole network that will prolong the lifetime of the network in an efficient manner. In this paper, a two-level fuzzy logic is proposed in choosing cluster head based on node localization and network traffic. In the upper decision making level called global level of qualification leads to better performance of the inference system based on all the above six fuzzy parameters for establishing an energy efficient network model. We develop an algorithm to calculate energy across the network if the source and destination is known. We evaluate the cost and benefit of the data fusion, in order to adaptively adjust whether fusion shall be performed for minimizing the total energy consumption when energy efficient node scheduling migrates from a particular node to another node. Simulation results show that EEDS gives the best performance with respect to network life time density and residual energy of the node.*

## 1. Introduction

There are many fundamental problems that sensor network research will have to address in order to improve optimal data delivery in remote monitoring environments. Efficient message routing techniques in a network can done by identifying clusters in the network model and then allocating cluster head for effective transmission of data. Numerous civil and disaster information gathering scenarios applications can be leveraged by networked sensors. In disaster management situations such as earthquakes, sensor networks can be used selectively map the affected regions directing emergency response units to take a quick action. The use of wireless sensor in hierarchial clustering algorithms in varied environment usage leads to many complications related to network lifetime. Wireless sensor networks and their physical environment is closely related to, and continue to change as the environment changes. These time-varying factors seriously affect the performance of the system and the network load and the dynamic change of the operating rules of the susceptible to changes in environmental factors external stimulus, such as low-power wireless communication; With the consumption of energy, the change of the system working state require sensor network system has the adaptability to dynamic environmental changes. Wireless sensor network consists of low cost nodes with limited power applications for information gathering in harsh terrains for weather and climate monitoring. An important characteristic of these nodes is that the nodes remain unattended and are resource constrained due to their deployment to some remote location. Due to such limitations of unpredictable behavior of nodes [1] it is necessary to implement optimum procedures that make the sensor nodes conserve energy to increase the lifetime of the network [2], [3], [4] for information processing and proper retrieval. The harmless P-waves are almost twice as fast as the S-waves, which cause most of the destructive shaking [5]. They also may include other parts of the application, such as positioning systems, power systems, and so on. By means of a built-in variety of sensors can measure temperature, humidity, barometric pressure, chemical and other physical phenomena. Multiple routes can be useful to compensate for the dynamic and unpredictable nature of ad hoc networks, also in energy and bandwidth constrained sensor networks. load balancing, fault tolerance, bandwidth aggregation, and reduced delay. Due to such limitations of unpredictable behavior of nodes [1] it is necessary to implement optimum procedures that make the sensor nodes conserve energy to increase the lifetime of the network [2], [3], [4] for information processing and proper retrieval. The harmless P-waves are almost twice as fast as the S-waves, which cause most of the destructive shaking [5]. They also may include other parts of the application, such as positioning systems, power systems, and so on. By means of a built-in variety of sensors can measure temperature, humidity, barometric pressure, chemical and other physical phenomena. Multiple routes can be useful to compensate for the dynamic and unpredictable nature of ad hoc networks, also in energy and bandwidth constrained sensor networks. load balancing, fault tolerance, bandwidth aggregation, and reduced delay. In order to deal with this issues , fuzzy logic algorithm is proposed. The fuzzy-logic algorithm is expected to be time-consuming due to the computational response time. However actual response time is dependant on the geographical distance which has been reduced using adaptive data fusion routing algorithm in wireless sensor networks. Multi-path routing protocols can increase the degree of fault tolerance by having redundant information routed to the destination over alternate paths. This increases the energy overhead, but helps to reduce the probability that communication is disrupted and data is lost in case of link failures. Overhead and integration of energy savings through fuzzy rule based analysis will result in energy gain, dynamic decision in node data fusion operation will provide the optimal energy consumption.

An Earthquake Early Warning System would require a smaller energy consumption as sensor nodes carry limited and not easy to replace the features require localization algorithm should be low-power energy. A high positioning accuracy which is an important indicator of the general measure of localization algorithm to calculate the ratio of error with wireless range, 20% said that the positioning error equivalent to 20% of the nodes in the wireless range. Calculation is distributed that compute nodes position completed node local distributed algorithm can be applied to large-scale sensor networks. Each node maintains an energy cost estimate for each of its path entries. This cost estimate determines the probability that a packet is routed over a certain path. If a node aims to transmit a packet to a certain destination for which it has multiple paths, it chooses the forwarding node according to a probability assigned to that path. Each intermediate node does the same and forwards packets according to the probability assigned to the different paths in the table. This is continued until the data packet reaches the destination node. Using this simple mechanism to send traffic over different routes helps in using the nodes' resources more equally. An overall gain of ~40% of network lifetime increase with this probabilistic routing scheme has been achieved. Taking suboptimal paths occasionally into account pays off as nodes use their scarce resources more equally, which helps to remove load from central forwarder nodes that would otherwise run out of energy first. the current entry for this destination is deleted and the packet source is taken as new next node towards the destination node. In order to balance the energy consumption for elongating the lifetime of this WSN, cluster identification among sensor nodes is given the top most priority. Each cluster would have a leader, often referred as cluster head (CH)[6-11].The CH is responsible for not only the general request but also receiving the sensed data of other sensor nodes in the same cluster and routing (transmitting) these data to the sink. Recently a lot of research efforts has been targeted in the aspect of optimization of communication for heterogeneous networks through design of routing protocols and fuzzy inference algorithms. So far in literature, many energy efficient routing and clustering algorithms are proposed in which a cluster head allocation is done that gives a greater robustness to transmit and forward sensing data to the base station [12]. LEACH [13], CHEF [14] and Gupta[15] are the important cluster head selection algorithms. But in most of the algorithms cumbersome inputs and cluster overhead scenarios negatively affects the network lifetime and also increases the data delivery latency in the network. This may prove costly in earthquake warning message transmission whereby a mini delay in data transfer might collapse the entire system. LEACH (Low Energy Adaptive Clustering Hierarchy) routing protocol with a group of clusters in broad region for analysis in an earthquake related information processing network may become incumbersome, as the cluster head uses a single hop routing method directly transmitting data from sensor nodes in the cluster to the base station for analysis. The distance between the cluster head and the base station growing larger than radio region of availability subsequently effects the local energy consumption for a certain threshold. The saving of a human life at an earthquake site, when using a sensor node loaded with LEACH becomes impractical as the node's range is increased due to a secondary shock of the earthquake. The LEACH and CHEF schemes have the same characteristic single hop network model. As a result, the system produce erratic communication overhead in multiple hop network models [16] during cluster-head selection process. In order to avoid the above scenario, the proposed study implements a two-way fuzzy logic approach to evaluate the qualification of sensors becoming a cluster head based on the nature of the network and network traffic. The two way local and global cluster head election provides modularity for the sensor groups utilizing intelligent technique is applied to improve the efficiency of wireless sensor network involved in the study. This algorithm allows a cluster head election based on unique criterion based on factors including available energy at a sensor node, number of neighboring sensor nodes, a node's distance from the current group leader and base station and characteristic metrics based on encounter level between nodes of the cluster in the current network. Using the lowest energy path for all packets is not necessarily best for the long-term health of a sensor network, as important forwarders might run out of energy first. However the local and global domain based inference analysis makes sure that the energy of the system prolongs over a certain time interval. The two- level inference mechanism provides modularity that gives parallelism and scalability for increasing energy efficiency of the clustering techniques involved in the study. Mathematical equations are framed to guide the selection process incorporating conditions that would increase the network lifetime as well. In local level the qualified nodes are selected based on their energy and number of neighbors of them that can be held trust worthy by its neighbors. Then, in the global level it is necessary to find the best node cooperation regarding to the average encounter relation metric. Appropriate evaluation based on cluster-head node election based on encounter of wireless nodes in a dynamic framework can drastically reduce the energy consumption and enhance the lifetime of the network. In this work, we evaluate EEDS,F3N[17] and LEACH by many simulation results through matlab and compared the performance to find the energy dissipation outputs in 500 rounds for energy routing. It is found that the fuzzy based algorithm EEDS considering characteristic metrics in network traffic performs better than F3N as shown through simulation. The rest of this paper is organized as follows. In Section 2, we discuss the related work for cluster head election. In Section 3, we introduce the proposed system and define the fuzzy rule base. In Section 4, we present the simulation results. Conclusions are given in Section 5.

## 2. Related Work

There are many fundamental problems that sensor network research will have to address in order to improve optimal data delivery in sensor networks. Efficient message routing techniques in a network can done by identifying clusters in the network model and then allocating cluster head for effective transmission of data. Energy and distance is a basic criterion in electing cluster heads. An analytic approach to choose cluster head out of N nodes with probability P leads to the selection of cluster heads in fig 1.By specifying a scenario where the red colored cluster head nodes blue colored nodes which are not in range and do not making any edge with header yields cluster head selection using a deterministic component for analysis as studied in [34].

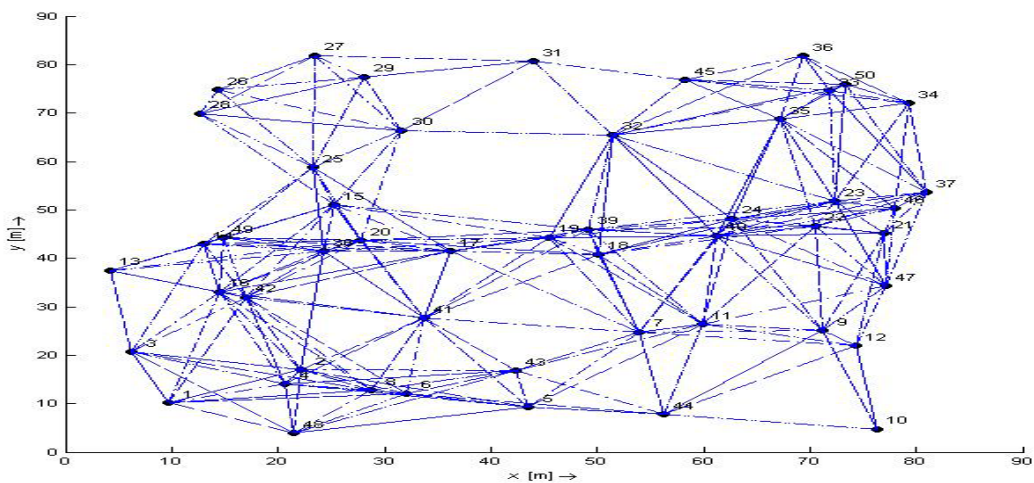

Fig 1: Network selection by edge based deterministic component

Selection of favorable cluster-head in earlier rounds result in an un-homogenous cluster-head routing in later rounds since LEACH tries to distribute energy consumption among all nodes. In order to make an energy efficient model for increasing network lifetime a deterministic cluster-head analysis outclasses a stochastic algorithm.In the proposed work a fuzzy inference technique provide more robustness in efficient cluster head allocation by allowing usage of decision parameters in network traffic. Implementation of network routing algorithm has been done in the proposed study by analyzing six fuzzy variables analyzed for cluster head selection through analysis of parameters of node's energy, node's traffic pattern and node's centrality for faster message transfer path in cluster head allocation scenarios. Earthquake warning systems require a very effective network model where civilians need to receive warning messages instantly. The delay due to traffic can be reduced by minimum message passing over de-congested network paths and proper cluster head allocation that can give better conflict control outputs. Fuzzy inference algorithms for cluster head allocation in a network has the advantage to reduce conflict for network traffic schemes between the network parameters that control clusters through making decision making rules. However the fuzzy inference schemes may slightly delay process of message transmission as it avoids any coupling between the offline data parameters extracted from the network clusters traffic. In the proposed approach, there is only one selected CH for each round, whereas more CHs are needed for balancing energy consumption and improving network lifetime. The cluster head collects data from nodes within the cluster and aggregates the data and report the aggregated information to the base station. By making only the cluster head communicate with the base station, the overheads occurring if all nodes in that cluster communicate with the base station would be reduced. Several significant routing algorithms has been compared as no single routing protocol can guarantee the efficiency in al1 network situations. Each node maintains an energy cost estimate for each of its path entries. This cost estimate determines the probability that a packet is routed over a certain path. If a node aims to transmit a packet to a certain destination for which it has multiple paths, it chooses the forwarding node according to a probability assigned to that path. Each intermediate node does the same and forwards packets according to the probability assigned to the different paths in the table. This is continued until the data packet reaches the destination node. Using this simple mechanism to send traffic over different routes helps in using the nodes' resources more equally. In LEACH, the nodes organize themselves into local clusters, with one node acting as the cluster-head. All non-cluster-head nodes must transmit their data to the cluster-head, while the cluster-head node must receive data. LEACH routing protocol model assumes that the sensor network nodes have the same energy. The distributed wireless sensor net work and the technology of wireless communication could realize the subsidence monitoring and measuring in remote deployments. One of the most important research is to design a network taking the net work communication capacity to be increased and the redundant data transmission reduced as the target for improved calculation method, an improved layer for routing needs to be validated. However in a heterogeneous network nodes are found to have variable energy and network traffic varying from cluster to cluster: decision making rules are needed to be formulated based on fuzzy inference engine f using the F3N technique and the Gupta method that minimizes the energy consumption by switching energy effective routing over LEACH. Gupta and Halgamuge [18] use fuzzy reasoning for deciding the best cluster-heads in a WSN. The former use three features to guide the choice: node concentration, energy level, and centrality. The formers use of energy measures and a fuzzy clustering algorithm gives better results than that of a subtractive clustering technique [19]. Beside the above mentioned inferencing algorithms, FRCA[20] uses the Fuzzy Relevance Degree (FRD) with fuzzy value to perform and manage clustering. CHEF uses the remaining available energy and distance as selection factors for more than one cluster head selection locally in each round. The fuzzy set includes nodes' energy and their local distances. CHEF generates a random number for each sensor and if it is less than a predefined threshold, $P_{opt}$, then the node's chance is determined.

The proper implementation of fuzzy decision making scenarios is based on node localization characteristics and encounter between nodes to propose the dynamics of the network. Most algorithms elect leaders by associating weights or use a iterative cost function or heuristic to generate minimum number of clusters. The input fuzzy variables transmission energy, remaining energy, rate of energy consumption, queue size, distance from centroid and proximity to base station. The fuzzy input variables have been defined in Figure The rule base therefore consists of $2^4 \times 3^2 = 144$ rules. There is a single output fuzzy variable, cost[21] that gives the defuzzified value of cost of link between two sensor nodes based on the energy aware routing condition. Figure 3 gives details of the input fuzzy variables for determining cost between sensor nodes for analysis of energy conservation, delay optimization and network based metrics. In determining the cost of link from node i to node j, "transmission energy" represents the energy needed to transmit a data packet from node i to j. Lower value of transmission energy leads to lower link cost. "Remaining energy" indicates the energy level of node j. Nodes with less value of remaining energy should be avoided in being selected as next-hop. The head aggregates the collected data and then, send these data directly to the sink. The iterative procedure used by sensor nodes by sending data to the head, the head aggregated the collected data and then, sent these data directly to the sink is termed as a round. Consequently, its lower value results in a higher link cost. "Energy consumption rate" of node j is another important parameter. It is possible to have a node with a high value of initial energy, resulting in a higher value of remaining energy in spite of its high rate of energy consumption. Nodes with high rate of energy consumption are, therefore, assigned higher link costs. The fuzzy input variable "distance from the centroid" enables selection of routes with minimum hops as the centroid election procedure plays the most significant role in cluster head selection strategy. The analysis of network traffic based on packet drop can be studied based on inter-user interactions in wireless networks. The cluster head election scenario analyzed the effect of traffic for the nodes through queue length of the packets and used distance from cluster analysis for nodes nearer to the centroid as the one with lower link cost. We found using the energy consumption rate as shown in fig 10 for EEDS, F3N and LEACH, the selection surface and the parameters defined for our system was better than the fuzzy inference engine F3N. F3N implements a fuzzy inference scheme based on the cluster head election scenario with the neighboring nodes[22,23] and the network traffic pattern. This makes the system more efficient allowing better selection of cluster heads for better energy utilization and fairer distribution for shorter path analysis.

## 3. Proposed Fuzzy Inference System

This paper focuses on how to resolve power conservation[24-28] in WSN by developing a novel approach of cluster-head selection using a fuzzy inference engine. Fuzzy logic has potential for dealing with conflicting situations and imprecision in data using heuristic human reasoning without needing complex mathematical modeling. As the non-linear characteristic of the nodes and the transport delay are not captured by the simulation model, the experimental results are an indication of the fuzzy controller's ability to handle modeling uncertainties. This change gave rise to a system with slower dynamics than the simulation model. Clearly, the type-2 FLC outperforms its type-1 counterpart. The main advantage of the

type-2 FLC appears to be its ability to eliminate persistent oscillations in centroid localization in the cluster head election based scenario. As the type-2 FLC can tolerate bigger modeling errors, it is more robust than type-1 fuzzy logic controllers.

### 3.1 Comparitive analysis of Fuzzy Inference algorithms

The proposed EEDS protocol successfully finds the stable region by being aware of heterogeneity through assigning probabilities of cluster-head election weighted by the relative initial energy of nodes. Simulation results of energy efficient routing of LEACH, F3N and EEDS protocol has been found to rate algorithm that takes the least energy for processing. We carried out the simulations for 500 rounds. From the simulation results, it was found that at the local level of fuzzy inference analysis, the probability of a sensor node to be a CH is increased with increase of number of neighbor nodes and remained battery power and is decreased with the increase of distance from the cluster centroid. Cell also has a separate state parameters (so-called ' records' ) record that they have non-zero state before the neighbors.

Between 0 and 1 generate a random number r. If the sum of> 0 (at least one neighbor) and r> threshold, or the cell has never had a neighbor, then cell = 1.If the sum> 0 set the " record" sign, record the cell has a non-zero neighbors.

i) F3N method: Cluster formation and cluster head scenario analysis was recently explored using fuzzy inference algorithm by comparing the performance of the proposed system for tree different parameters: residual power of nodes, degree of number of neighbor nodes (D3N); and Distance from Cluster Centroid (DCC). The decision making procedures access a sensor with sufficient battery and computational power without analyzing data latency and energy-aware routing decisions affects the decision making capabilities of the system. A set of wireless nodes in a cluster may have a potentially higher value for initial energy, resulting in a higher value of remaining energy in spite of its high rate of energy consumption. Neighbor nodes selection may be affected depending on the position as the centroid as shown in figure 1 that tries to cover as much information geometrically without prior effecting the redundancy penalty for the site.

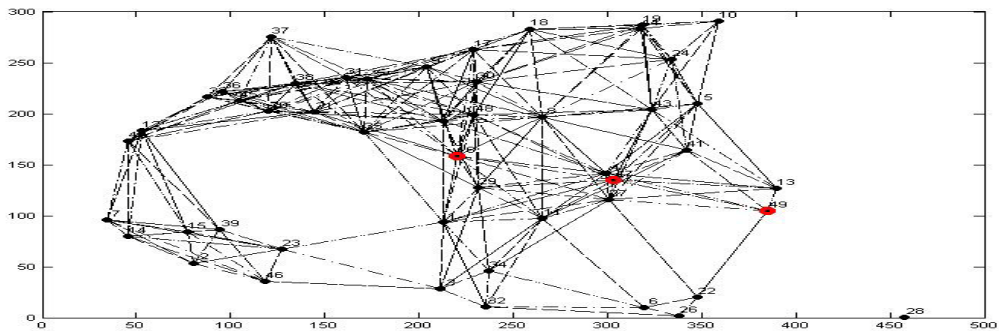

Fig 2:Centroid selection with node localization in a cluster head election scenario

ii)EEDS: In order to avoid the above scenario where the number of neighboring nodes may affect the centroid selection for a system that is a limitation in F3N the paper transforms clustering and cluster head selection scenario using a two-level fuzzy logic based routing algorithm. The Energy Efficient Dynamic Scenario(EEDS) scheme uses fuzzy logic controller (FLC) where metrics are assigned at the local and the global level are shown in Fig. 2 and fig 3 respectively. The blocks defined include a fuzzifier, inference engine, fuzzy rule base (FRB) and defuzzifier. As membership functions, we use triangular and trapezoidal membership functions because they are suitable for real-time operation [29] a two level fuzzy logic[30,31] approach to evaluate the qualification of sensors to become a cluster head .The cluster head election is proposed based on six fuzzy descriptors – transmission and residual energy, energy consumption rate , queue size, centrality of cluster head allocation and proximity to the base station to assign a dynamic priority for network transmission schemes in wireless sensor networks.

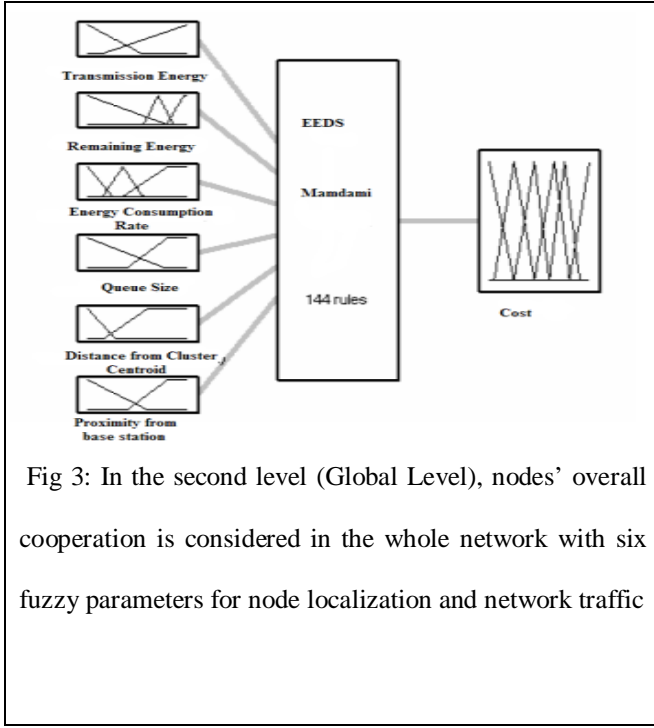

Fig 3: In the second level (Global Level), nodes' overall cooperation is considered in the whole network with six fuzzy parameters for node localization and network traffic

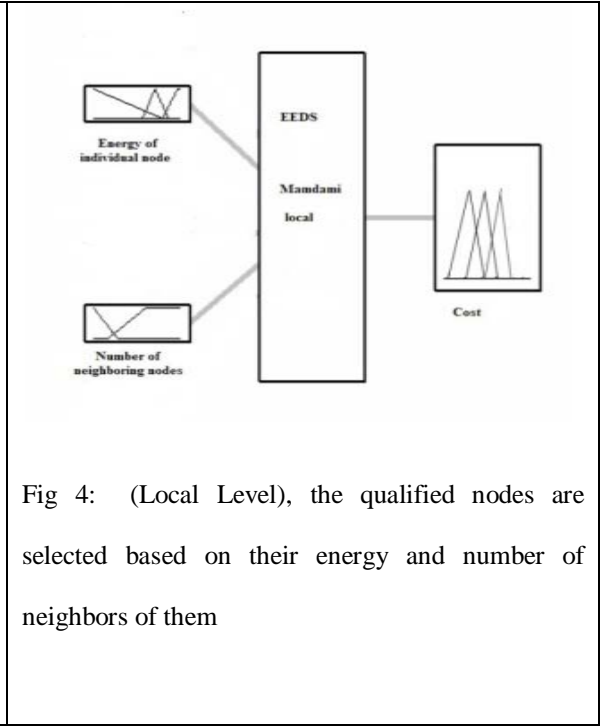

Fig 4: (Local Level), the qualified nodes are selected based on their energy and number of neighbors of them

MIN-MAX inference technique has been used in the fuzzy controller. There is a single output fuzzy variable, namely Cost, the defuzzified value of which determines the Cost of link between two sensor nodes. To find a crisp output value from a solution fuzzy region, the controller uses Centroid Defuzzification method. The membership functions for input parameters of FLC are defined as:

T (tr_energy) = {Low($L_o$),High($H_g$)};
T(r_energy)={Low,Medium,High};
T(q_size)={small,large};
T(d_centroid)={small, large};
T (e_rate) = {Small($F_w$),Medium($M_e$),High($H_g$)};
T (proximity) = {small,large}.

This might be in conflict with the layered design paradigm, but in wireless sensor networks with scarce energy resources, such cross-layer approaches are acceptable, if higher efficiency can be achieved. increased the probability that transmissions along multiple paths interfere with each other. Load balancing of multipath-routing was deteriorated by the additional cost of coping with path interference. Each node maintains an energy cost estimate for each of its path entries. This cost estimate determines the probability that a packet is routed over a certain path. If a node aims to transmit a packet to a certain destination for which it has multiple paths, it chooses the forwarding node according to a probability assigned to that path. Each intermediate node does the same and forwards packets according to the probability assigned to the different paths in the table. Methods developed with fuzzy logic propositions have been shown to be useful in difficult conditions with respect to non-linear and time-variant systems. When dealing with complex methodologies varying with time, fuzzy logic based methods can be used for faster convergence and reduced complexity with a slight degradation in performance compared to that of standard methods. We have used the most commonly used fuzzy inference technique called Mamdani Method [32] due to its simplicity. The process is performed in four steps: Fuzzification of the input variables energy, concentration and centrality and network traffic by taking crisp inputs from each of these metric output to determine the degree to which these inputs belong to each of the appropriate fuzzy sets. Consequently, the routes are selected so as to avoid nodes with low remaining energy, thereby extending the lifetime of the sensor network selecting of the cluster head is not easy in different environments as different wireless nodes have robust yet varied interaction event among MNs termed as encounters, defined as the event when two MNs move into the radio range and they are able to communicate with each other directly if the users choose to do so. In each cluster, termed as Ci,j, one of sensor nodes in Ci,j is elected as the coordinator Hi,j determining queue size for nodal encounter analysis. The coordinator Hi,j analyzes the sensed data of a sensor, without aggregating other data, sends the data to the head Hi and the head Hi then send the data to the sink. The selection of the coordinator is done based on three metrics The clustering coefficient (CC) is used to describe the tendency of nodes to form cliques in a graph. Clustering coefficient is the average ratio of neighbors of a given node that are also neighbors of one another. CC indicates higher tendency that neighbors of a given node are also neighbors to each other, or relationship between MNs formed through encounters. Average path length (PL) is used to describe the degree of separation of nodes in the ER graph. Disconnected ratio (DR) is used to describe the connectivity of the ER graph. CC and the PL for regular graphs and random graphs with the same corresponding total node number M and average node degree d. The output linguistic parameter is the Possibility of CH Selection (Cost). We define the term set of Cost function as: {Very Weak (VW), Weak (W), Little Weak (LW), Medium (MD), Little Medium(LM), Strong (S), Very Strong (VS)}. Encounters are important events in wireless networks as they provide chances for MNs to directly communicate, even without an infrastructure to the sink. More encounter means more loss of residual energy so more number of unique encounters means more loss of energy. Higher CC indicates higher tendency that neighbors of a given node are also neighbors to each other, or heavy "cliquishness" in the relationship between MNs formed through encounter for sinks. By removing encounters with short durations does not cause abrupt degradation in the performance of information diffusion, in terms of both the node range and the average delay. The objective of our fuzzy routine is to determine the value of Cost for a link between two sensor nodes such that the life of a sensor network is maximized. The lifetime of wireless sensor networks is generally defined as the time when the energy level of the first sensor node becomes zero. The fuzzy rule base as shown in Figure 4 displays our fuzzy model.

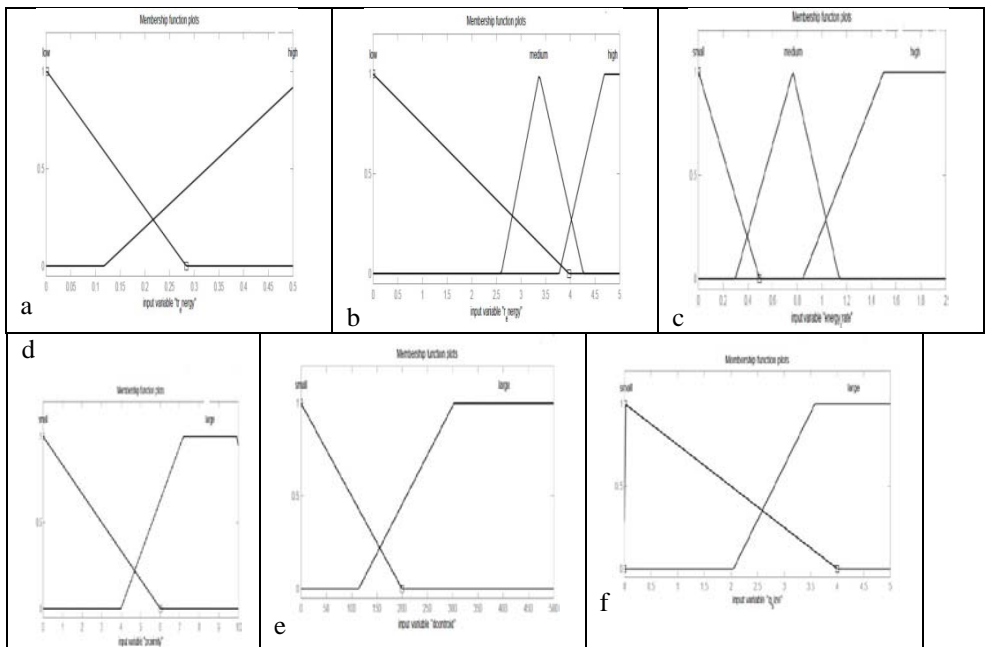

Fig 5: Input fuzzy variables for a)transmission energy, b)remaining energy, c)rate of energy consumption, d)proximity from base station, e) distance from centroid and f)queue size for nodal encounter analysis.

The membership functions for the output parameter Cost are defined as as in figure 5.
µVL(COST) = g(COST;VL0, VL1, VLw0, VLw1);
µL(COST) = f(COST;L0,Lw0,Lw1);
µLM (COST) = f(COST;LM0, LMw0, LMw1);
µHM(COST) = f(COST;HM0,HMw0,HMw1);
µH(COST) = f(COST; H0, Hw0, Hw1);
µVH(COST) = f(COST; VH0, VHw0, VHw1).

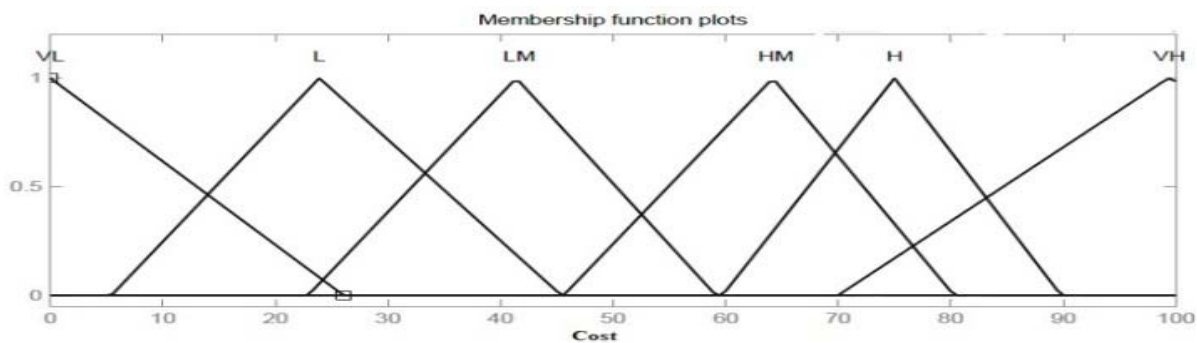

Figure 6: Output fuzzy variable for Cost having VL(very low),(L)Low, LM(Low Medium),HM(High Medium),H(High) and VH(Very High) cost path

## 4. Simulation and results

In this section, we evaluate the performance of our proposed algorithm in MATLAB. In four different scenarios, 30 and 60 nodes are randomly distributed in a 60×60 and 120×120 m2 network. The initial energy of the sensors is 0.1J. Simulation was run for 500 rounds. We use a simplified model proposed in [8] for the radio hardware energy dissipation. Proximity to BS parameter is also measured by total distance from cluster heads to the BS. In addition, the BS is capable of measuring the distances among all CH. In the global level these three parameters are considered as the inputs to the fuzzy system and sensors' qualification parameter for becoming a CH are considered as the output. The fairness of energy consumption can be well observed by measuring the variance of the residual energy of all nodes in each round. For comparison with the crisp approach, the same scenario was simulated using non-fuzzy/crisp variables. In LEACH the data is chosen according to the position of the sink. A sensor node in the cluster is elected randomly as the cluster head. The link Cost is again determined on the basis of percentage of remaining energy of each node. Consequently, the routes are selected so as to avoid nodes with low remaining energy, thereby extending the lifetime of the sensor network. Residual Energy in each round is a good metric to measure the energy efficiency. We compare our proposed algorithm EEDS to LEACH and F3N network's residual energy distribution over 500 rounds. The less variant residual energy in each consequent round is the reason of the fairer energy consumption. Contrary, more variance in energy consumption shows that the network's load is on some sets of nodes. This nodes which play the role of cluster head selection to WSN to improve energy efficiency based on encounters and cost of path between nodes. It presents a comparison between the different methods on the basis of the network lifetime. It proposes a modified approach for cluster head selection with good performance and reduced computational complexity.

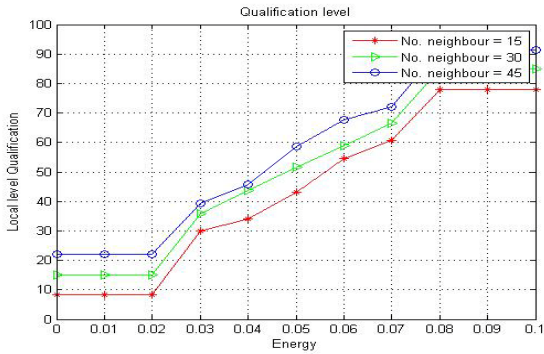

Fig 7:Local qualification level for varying node energy

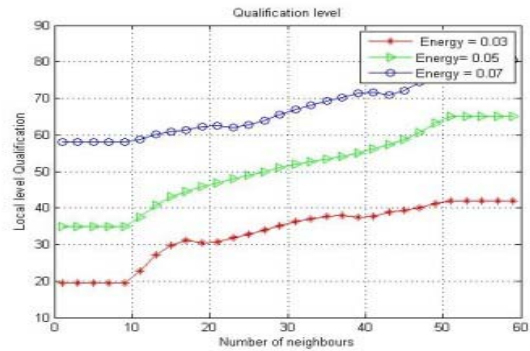

Fig 8:Local qualification level for varying node neighbor

In wireless sensor networks, the focus of multi-path routing is often on load-balancing or fault tolerance, rather than on the aggregation of bandwidth. Often, the goal of multi-path routing protocols is maximize the time the network is operable and fulfills its observation task. Using multiple paths in ad hoc networks to achieve higher bandwidth, balance load or achieve fault tolerance is not as easy as in wired networks. Power consumption is one of the most influential factor in the design and implementation of WSN. Because of the limited battery power of the sensor nodes, under the premise of ensuring the positioning accuracy of the amount of computation required by the closely related and power positioning communication overhead storage overhead, the time complexity is a group of key indicators.

Residual energy of network: Residual energy of network in each round can be a good metric to measure the energy efficiency of the algorithms and by viewing and saving the residual energy of each node the number of surviving .

Distance among nodes: One of the important parameters in clustering is the distance among nodes which has a greater effect on the degree of centrality which has a direct effect on the distribution of nodes and election of cluster head. Suitable distribution of CHs results in more balanced load to be on the nodes. We evaluated EEDS,F3N and LEACH ability over 500 rounds as shown in figure 9. From the simulation results, we found that the probability of a sensor node to be a CH or the cost is different for different scenarios of the algorithms. The energy consumption increased initially increased with increase of number of neighbor nodes and remained battery power. LEACH performs the least effectively among the three algorithms. Although F3N performs as good as EEDS in the initial levels EEDS does better inferencing as the algorithm can also make decisions based on the queue size having a direct impact on the network traffic rate. The requirement is to establish energy consumption would be to employ a dynamic cluster formation scheme. All the three algorithms however perform in a stable manner over the 500 rounds of simulation as in fig 9.

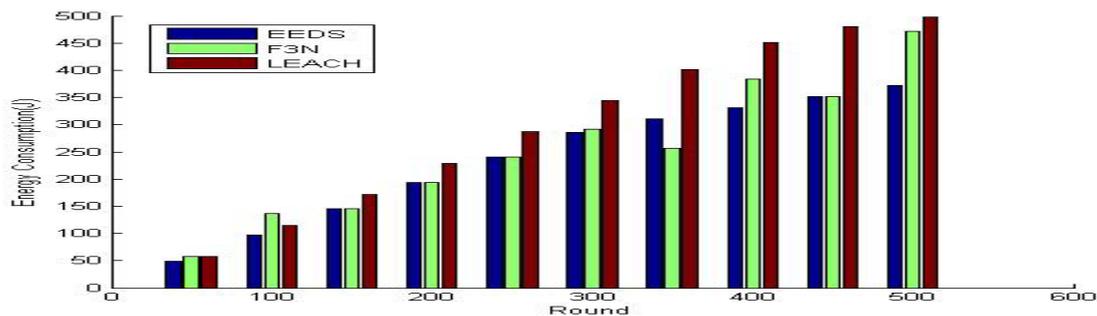

Fig 9: Comparative analysis of energy consumption with various rounds for EEDS, F3N and LEACH

The network lifetime can be improved if the clustering algorithms are made distributed as in LEACH as shown in fig 9. In all of the methods, the energy parameter is taken into consideration only during cluster head selection. The performance may be increased by considering energy as a parameter during clustering itself. Nodes are identified by identifiers which are unique and totally ordered. Each node maintains a routing table which entries point toward that node's immediate predecessors and $r/2$ immediate successors an integer value, a priori known by all nodes). Forwarding entries to their successors and predecessors along multi-hop physical paths. As a result, a node contains routing table entries that point toward its nearest neighbors in ID space, but also entries for the paths between other pairs.

## 5. Conclusions

We proposed a new fuzzy based clustering multi-hop protocol, EEDS which ensures that the cluster heads are selected with plausible transmission energy, dynamically changed and the heads are well connected based on the encounters. The study shows that by choosing a node with more residual energy aids in optimal energy consumption to extend the network lifetime. However, only considering the residual energy may lead to a waste of network energy. With the increase of the network traffic and the increase of the distance between the sensor and sink, the possibility of a sensor to be selected as CH is decreased. Conserving power prolongs the lifetime of a node and also the lifetime of the whole network. Clustering is one of the energy-efficient techniques for extending the lifetime and scalability of a sensor network and dynamic network analysis for encounter between nodes. Probability of a sensor node to be a CH is increased with increase of number of neighbor nodes and remained battery power and is decreased with the increase of distance from the cluster centroid . With the introduction of the forced cluster head based on fuzzy inference rules for finding data relativity of the monitoring stations and measuring variation for the unit pitch points within the continued time, the redundant data latency can be brought down. A comparison simulation on the network for existing time of the improved algorithm for the LEACH and fuzzy rule base protocols was conducted comparing EEDS with F3N and power constrained LEACH. We found that EEDS performs better than all other protocols. Moreover, a robust load distribution and a variance in energy consumption process for the two step fuzzy approach shows that it can be easily tuned for different network and node conditions simply by changing shapes of the fuzzy sets that will increase the efficiency of the proposed algorithm. A topic for future research is the analytical study of implementing fuzzy logic implementation for variables that allow the cluster heads to be more sparsely distributed [33,34].